# Influence of the pattern shape on the efficiency of front-side periodically patterned ultrathin crystalline silicon solar cells

Aline Herman,[1] Christos Trompoukis,[2,3] Valérie Depauw,[2] Ounsi El Daif,[2] and Olivier Deparis[1]

[1]*Solid-State Physics Laboratory, Facultés Universitaires Notre-Dame de la Paix (FUNDP), rue de Bruxelles 61, B-5000 Namur, Belgium*

[2]*Imec, Kapeldreef 75, B-3001 Leuven, Belgium*

[3]*Departement Elektrotechniek – ESAT, Katholieke Universiteit Leuven, Kasteelpark Arenberg 10, B-3001 Heverlee (Leuven), Belgium*

Patterning the front side of an ultrathin crystalline silicon (c-Si) solar cell helps keeping the energy conversion efficiency high by compensating for the light absorption losses. A super-Gaussian mathematical expression was used in order to encompass a large variety of nanopattern shapes and to study their influence on the optical performance. We prove that the enhancement in the maximum achievable photo-current is due to both impedance matching condition at short wavelengths and to the wave nature of light at longer wavelengths. We show that the optimal mathematical shape and parameters of the pattern depend on the c-Si thickness. An optimal shape comes with a broad optimal parameter zone where experimental inaccuracies have much less influence on the efficiency. We prove that cylinders are not the best suited shape. To compare our model with a real slab, we fabricated a nanopatterned c-Si slab via Nano Imprint Lithography.

## I. INTRODUCTION

In order to compete with other renewable energy technologies, the price of photovoltaic (PV) electricity has to be reduced by at least a factor of two.[1,2] Solar cell technology based on crystalline silicon (c-Si) presently dominates PV, attaining more than 80% of the market, and is expected to stay at a very high level during the next years.[3] There are various reasons for this: c-Si absorbs a large range of the solar radiation spectrum, it is very abundant, non-toxic, stable and can be processed using well known techniques.[4] However, the cost of present c-Si technologies is still high (a silicon wafer represents 30% of a PV module cost). On the other hand, low cost thin film PV technologies have been developed in the last 30 years. However, the sustainability of their development will be limited due to the rareness or the toxicity of the elements (CdTe and CuIn(Ga)Se) or to their low efficiency (e.g. CdTe, amorphous silicon).[5,6]

The use of ultrathin c-Si films, with thicknesses of a few microns may potentially overcome these two limitations[7]. The term ultrathin is justified not only by the fact that the film thickness is much lower than the c-Si absorption length but also by the technical challenges related to the fabrication of micron- or even sub-micron size films of c-Si. This is allowed by the development of cheap methods to fabricate thin crystalline silicon. This approach could help reducing further the PV costs.[8,9]

However, this approach also brings new challenges, and until now the cell efficiencies have not yet reached their high potential. In particular, the radical thickness reduction results in a significant reduction of the absorption in the near-infrared region of the solar spectrum due to the indirect band-gap of c-Si.[10] So, ultrathin c-Si PV could potentially replace the currently dominant Si-wafer-based technology provided that solutions are found to keep solar light absorption high. One solution to improve the light absorption efficiency is to use front-side or/and back-side surface texturing which is already well-known to help coupling incident light into the active layer via light trapping techniques.[11-13]

As a way to improve light harvesting in ultrathin (wavelength scale) crystalline silicon slabs, we pattern the front-side with a periodic square array of holes at the incident light wavelength-scale (improving absorption at infrared wavelengths for which light trapping is needed).[14-17] Some studies on specific pattern shapes (nanodomes[18], triangles[19], honeycomb[20-22], inverted-cones[23], inverted-pramids[24,25], pyramids[26]) already exist. A general study of the antireflective mechanism for different pattern profiles has also been done[27]. We propose to use a generic mathematical function (super-Gaussian) for describing a large range of very different shapes of holes, hence surface pattern morphologies. This methodology could be helpful for the design and fabrication of new ultrathin c-Si solar cells consuming less quantity of material. Indeed, nowadays, front-side texturing techniques (such as random pyramid texturing) consume a lot of photoactive material.

## II. SURFACE MORPHOLOGY MODELS

To increase the absorption of ultrathin c-Si slabs, we pattern the front-side with a periodic square array of holes at the wavelength scale. We focus on the theoretical case of a stand-alone slab of c-Si in air, with the idea to isolate the influence of surface morphology on the absorption avoiding the influence of other layers (e.g. metallic back reflector, anti-reflection coating). In the presence of additional layers which can be used as substrates, the optical response of the whole stack will be different. Generally, a material with a refractive index different from air at the slab back side will result in a different impact on light impinging at the back side. We chose therefore to work on an (hypothetical) stand-alone c-Si slab in order to understand the optical effects in the c-Si. When it will come to fabricate cells, we will consider a more realistic multilayer taking the whole cell structure into account, which however would not bring more insight on the influence of the surface morphology.[16,28,29]

Our experience in 3D optical simulations of biological photonic structures told us that the use of a hexagonal array would not change significantly the outcomes of the present study.[30] Therefore, we choose a periodic square array of three-dimensional (3D) holes of various shapes. We describe the mathematical shape of the holes using a normalized super-Gaussian profile function, i.e. Eq. (1).

$$f(r) = \exp\left[-\left(r^2/(2\sigma^2)\right)^m\right] \qquad (1)$$

where $f(r)$ represents the value of the super-Gaussian profile function at a given radius $r$ and $m$ is a real number which determines the actual shape of the holes. From eq. (1), we define the altitude $Z(r) = B/f(r)$, where $Z$ is counted from the bottom of the slab (Fig. 1), $B=T-h$, $T$ is the thickness of the slab, $h$ is the height of the hole. When $m<1$ or $m>1$, the hole has a convex or concave shape, respectively. If $m \to \infty$ then the hole becomes a cylinder. $\sigma$ is a parameter depending on the radius $R$ of the hole on the top surface : $\sigma^2 = R^2/2\left[\ln(T/B)\right]^{1/m}$. The value of $\sigma$ is deduced from the definition of $Z(r)$ by imposing $Z(r=R)=T$ where $R$ is the radius at the top of the hole. All these parameters are presented on Fig. 1.

    The absorption spectrum of the patterned c-Si slab is calculated numerically using Rigorous Coupled-Wave Analysis (RCWA). As far as the implementation of RCWA is concerned, we use a 3D-scattering-matrix electromagnetic computation code where the slab has to be divided into layers. The assembly of these layers forms the complete patterned slab. Because of the use of such layers, the mathematical profile of the holes is discretized. Each hole is formed by a stack of air cylinders with decreasing radius from the top surface (Fig. 2(a)). The radius of the air cylinder embedded in layer #$l$ is represented by $r_l$. The radius is given by $r_l = \sqrt{2}\sigma\left[\ln\{Z_l/B\}\right]^{1/2m}$, where $Z_l$ is the mean depth of layer #$l$. The value of $r_l$ is deduced from the definition of $Z(r)$ by setting $r=r_l$ and $Z(r_l)=Z_l$. The thicknesses of the layers are calculated via a method leading to a computational shape as close as possible to the mathematical one. This method uses the wavelength of light as a criterion for discretizing the profile. In practice, we allow a maximum deviation from the real profile by a factor given by

$\lambda_{min}/10$ with, in our case, $\lambda_{min} = 300\ nm$.

The main advantage of using this mathematical super-Gaussian expression is that it allows us scanning a large range of shapes (concave (Fig. 2(b)), convex (Fig. 2(c-d)), cylindrical (Fig. 2(e)) etc.) by only changing the value of the parameter $m$.

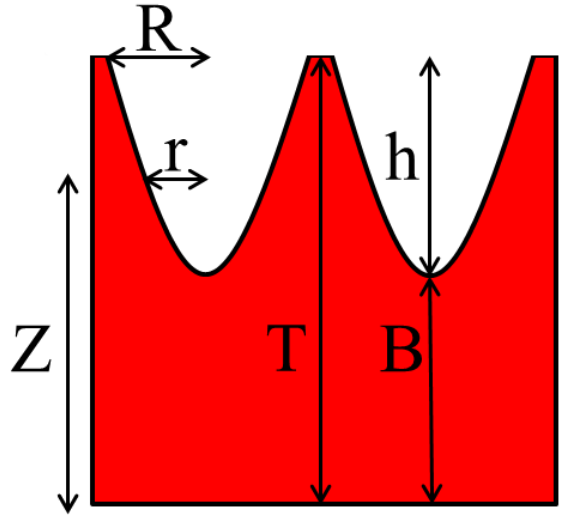

FIG. 1. (Color online) Parameters of the super-Gaussian function describing the morphology of the holes patterning the front surface.

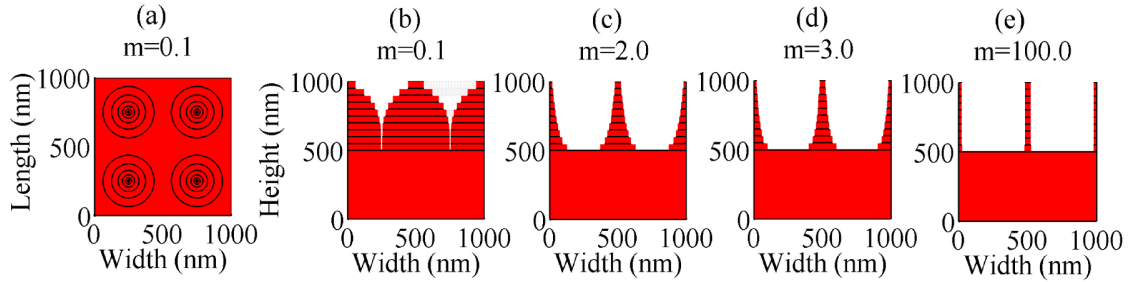

FIG. 2. (Color online) The top surface of the c-Si slab (red) is patterned with air holes (white). For the needs of the electromagnetic calculation, the slab is divided into layers with different thicknesses. Each layer contains an air cylinder. The stacking of the different cylinders forms the air hole. A large range of shapes is available thanks to the use of a super-Gaussian function: (a) top view (m=0.1), (b) cross section view (m=0.1); (c) (m=2.0); (d) (m=3.0) and (e) (m=100.0).

As we wanted to identify the best parameters, maximizing the light absorption in the patterned c-Si slab, we performed several calculations varying the parameters of the super-Gaussian function. For each silicon thickness studied, the thickness of the slab was kept constant while varying the hole parameters, in order to be consistent with reality; i.e. where the patterning is done after the fabrication of the silicon layer.

## III. METHODOLOGY

The goal of the study is to find the pattern parameters that maximize the light absorption and hence the photo-current in the active layer of a solar cell.

### A. Choice of the studied c-Si slabs

As mentioned above, we focus on the theoretical case of a c-Si slab in air. This choice allows us to take into account only the absorption in the c-Si. For that purpose, we used the wavelength-dependent crystalline silicon permittivity data from Palik's handbook.[32]

We consider several thicknesses of c-Si: an ultrathin slab (500 nm), a 1-µm-thin c-Si slab and a thick c-Si slab (700 µm). The thick slab absorbs the totality of the remaining part of the incident solar radiation that is not reflected at the air/c-Si interface, up to the band-edge of c-Si (≈1100 nm). Therefore, the effect of the back interface can be neglected in this case and this c-Si layer can hence be identified as a semi-infinite medium, avoiding back-reflection. In both 500-nm and 1-µm-thin slabs, on the other hand, back reflection and wave interference effects in general are of primary importance since their thickness is of the order of the relevant wavelengths.

### B. Effective medium theory

In order to study the specific contribution of impedance matching in the absorption enhancement, we use the effective medium theory, consisting in replacing an inhomogeneous layer by an effective homogeneous one. In the present case, the effective permittivity of each homogeneous layer is calculated by the Maxwell-Garnett[27] expression:

$$\varepsilon = \varepsilon_{c-Si} + 2f\varepsilon_{c-Si} \frac{\varepsilon_{air} - \varepsilon_{c-Si}}{\varepsilon_{air} + \varepsilon_{c-Si} - f(\varepsilon_{air} - \varepsilon_{c-Si})} \qquad (2)$$

where $f$ is the air filling fraction of the patterned c-Si structure, $\varepsilon_{air}$ is the permittivity of the "island" (the air hole in our case) and $\varepsilon_{c-si}$ is the permittivity of the host, i.e. substrate (c-Si).

We use the effective medium approach in two ways. In the first, the inhomogeneous part of the slab (c-Si/air holes) is replaced by a single homogeneous layer with an effective permittivity calculated from Eq. (2) using the air filling fraction in the patterned section of the slab. This approach was used as a starting point of comparison between our patterned structure and this simple case. In the second approach, each inhomogeneous layer containing a cylinder of air embedded into the c-Si host is replaced by a homogeneous layer. Their effective permittivity is simply calculated by Eq. (2). The first approach treats the whole pattern as a single effective layer whereas the second one describes it as a graded-index layer, giving an indication on the impedance matching.

In the structures under study, the distance between two consecutive inclusions (holes) is of the order of magnitude of the incident wavelengths. Therefore, it should be noted that we are not in the appropriate conditions for using any effective medium theory[33,34] (Bruggeman[35] or Maxwell-Garnett[27]) since it is only accurate when the incident wavelength is much larger than the distance between inclusions. Therefore, we use the effective medium theory in order to get a qualitative point of comparison with RCWA exact calculations.

## C. Quantities defining the optical performances

In order to characterize the optical performance of the nanopatterned c-Si slab, we use the integrated quantum efficiency defined by $\eta$, Eq. (3).[36,37] This quantity simply represents the percentage of incident photons that are absorbed in the c-Si. It depends on the incident (spectrally integrated from 300 nm to 1200 nm, i.e. c-Si absorption spectrum) photon flux: $\Phi_{inc} = \int \lambda/(hc) \times S(\lambda) d\lambda$ and on the absorbed (spectrally integrated) photon flux: $\Phi_A = \int \lambda/(hc) \times S(\lambda) A(\lambda) d\lambda$, where $h$ is the Planck constant, $c$ the speed of light, $\lambda$ the incident wavelength, $A(\lambda)$ the absorption spectrum and $S(\lambda)$ is the normalized solar spectrum AM1.5G:

$$\eta = \frac{\Phi_A}{\Phi_{inc}} = \frac{\int_{\lambda_{min}}^{\lambda_{max}} \frac{\lambda}{hc} S(\lambda) A(\lambda) d\lambda}{\int_{\lambda_{min}}^{\lambda_{max}} \frac{\lambda}{hc} S(\lambda) d\lambda}. \quad (3)$$

In order to characterize the electrical performance of the nanopatterned c-Si slab, we consider the ideal case where each absorbed photon is converted into an electron-hole pair (with no recombination i.e. perfect carrier collection, internal quantum efficiency equal to 1). We compare the maximum achievable photo-current in the active layer ($J_A = e\Phi_A$) with the solar harvested photo-current ($J_{inc} = e\Phi_{inc}$). The solar harvested photo-current assumes that we are in the ideal case of a perfect solar absorber where the entire solar spectrum is absorbed ($A(\lambda)=1$ for all $\lambda$).

We also define the Figure Of Merit of the photo-current ($FOM_J$), Eq. (4), as the ratio between the achievable photo-current in the nanopatterned slab ($J_A$) and the achievable photo-current in a reference ($J_R$):

$$FOM_J = \frac{J_A}{J_R} = \frac{\Phi_A}{\Phi_R}. \quad (4)$$

The reference is a planar homogeneous c-Si slab of the same thickness as the patterned slab.[16,38,39] However, we use a SiNx thin film as anti-reflection coating (ARC) since practical devices always use it for reducing reflection losses. The thickness of the ARC is optimized for each thickness of c-Si, in order to maximize the integrated quantum efficiency ($\eta$). Therefore, we used a 55-nm-thick ARC to the 500-nm-thick c-Si slab, (60-nm ARC to the 1-µm-thick slab and 80-nm ARC to the 700-nm-thick slab). We use a constant (wavelength independent) and real (no absorption) permittivity for the ARC layer. The refractive index value is chosen to be $n=2$ ($\varepsilon=4$) which is a good representative value of SiNx, no c-Si solar cell exists without ARC, once the optimal parameters of the patterned c-Si are determined, we perform simulations on the c-Si slab combined with a conformal anti-reflective coating on top of it.

### D. Validation of the models

In order to validate our model, we compare some experimental absorption spectra with the corresponding simulated ones. Two samples were investigated: a 1-µm-thick slab and a 700-µm-thick slab (bulk wafer). The 1-µm-thick slab was fabricated by an epitaxy-free (epifree) technique developed at Imec.[40] This epifree process provides high-quality silicon films without using epitaxy. It resulted in the formation of a high-quality monocrystalline ultrathin film (by the reorganization upon annealing of cylindrical macropores that can merge at high temperature and transform into a uniform plate-like micron-thin film). The samples were then patterned via Nano Imprint Lithography (NIL)[20,41-44]. The nanopattern was fabricated using a combination of soft-thermal nanoimprint lithography and reactive ion etching. Experimental details on the fabrication of the samples can be found in litterature[44]. The NIL technique offers a combination of high throughput, large area patterns with nanometer scale resolution and is therefore a potential solution for implementation in a photovoltaic production line.[20]

The samples were optically characterized using a Lambda 750S Perkin Elmer spectrophotometer. The measurements were performed in the 300-1200 nm wavelength range using an integrating sphere combined to a center mount (the sample was hold inside, at the center of the sphere). The center mount avoided performing two separate measurements (reflectance *R* and transmittance *T*) which could lead to inaccuracies due to sample positioning differences. The calculation of the absorption spectrum was carried out directly from the single measurement of the quantity $R+T$: $A=1-(R+T)$ from the energy conservation law.

## IV. RESULTS AND DISCUSSION

### A. Optimization of pattern parameters

Simulations were performed in order to optimize the period of the square array, the height, the diameter and the shape of the holes. We looked for the optimum values for different thicknesses of c-Si: ultrathin slabs (500 nm, 1 µm) and a thick slab (700 µm). The parameters ranges were chosen as follows. The period ($P$) was varied from 250 nm to 1250 nm (by a step of 250 nm). This choice was made because it corresponded to the range of wavelengths that could be absorbed by the c-Si. The diameter ($d$) of the holes was varied proportionally to the period. Instead of $d$, we used the ratio $\alpha=d/P$ as parameter, which was varied from *0.1* to *0.9* by steps of *0.1*. The maximum diameter was limited to $d=0.9P$ ($\alpha=0.9$) for a reason related to practical fabrication. It should be noted that the hole diameter is not restricted to be smaller than the period. The super Gaussian function provides profiles for $\alpha >1$ as well. However, taking a diameter greater than the period leads to a (virtual) waste of c-Si. Indeed, in this case, the top part of the slab is (virtually) removed (and therefore lost) because of the intersection between adjacent holes.

For the 1 µm and 700 µm slabs, the height ($h$) of the holes was varied from 50 nm to 700 nm. For the 500 nm, $h$ was varied from 25 nm to 350 nm. The last parameter was the shape of the holes. We considered extremely different shapes (convex, concave and close to cylindrical) by varying the parameter m from *0.1* to *100*.

In all the simulations, we used non-polarized light at normal incidence, where ($x, z$) plane was the plane of incidence (polar and azimuthal angles equal to *0°*). We tested the numerical convergence of the results prior to our simulation works. The use of *169* orders of diffraction (*13* plane waves along the $x$ direction and *13* plane waves along the $y$ direction, respectively, for the 2D Fourier expansions) turned out to be necessary to reach numerical convergence.

For all thicknesses of c-Si and for all parameters, we found that the optimal value of $\alpha$ was equal to *0.9*. This is not surprising since this value corresponds to a very small amount of planar silicon at the top surface and a smooth transition between air and silicon, with a progressive amount of silicon filling factor seen by light, allowing the graded index effect to take place. This conclusion was already reported in literature[27]. For this reason, we only present results with *α=0.9* (*d=0.9P*). In the following, we focus on the period of the square array, the height of the hole and its shape.

The first quantity we focus on is the integrated quantum efficiency ($\eta$). Fig. 3 shows maps of $\eta$ for three thicknesses (500 nm, 1 µm and 700 µm) and four different shapes (*m=0.1*, *m=2*, *m=3* and *m=100*). These maps reveal that the shape of the holes influences the maximal value of $\eta$. At first glance, we could think about neglecting the shape and thus use any type of hole. However, the shape of the holes also influences the size of the optimal parameter region.

For the 1-µm-thick c-Si slab, we notice the presence of two optimal zones for *m≥3*. These two zones are associated to a deep pattern (*h=700 nm* for *m=3*) and to a shallow one (*h=400 nm* for *m=3*), respectively (*h* values were identified from numerical calculations). The optimal patterns change with the value of *m* (*h=300 nm* and *h=600 nm* for *m=100*). The existence of deep and shallow configurations was already shown by Bozzola and co-workers for other pattern parameters.[15]

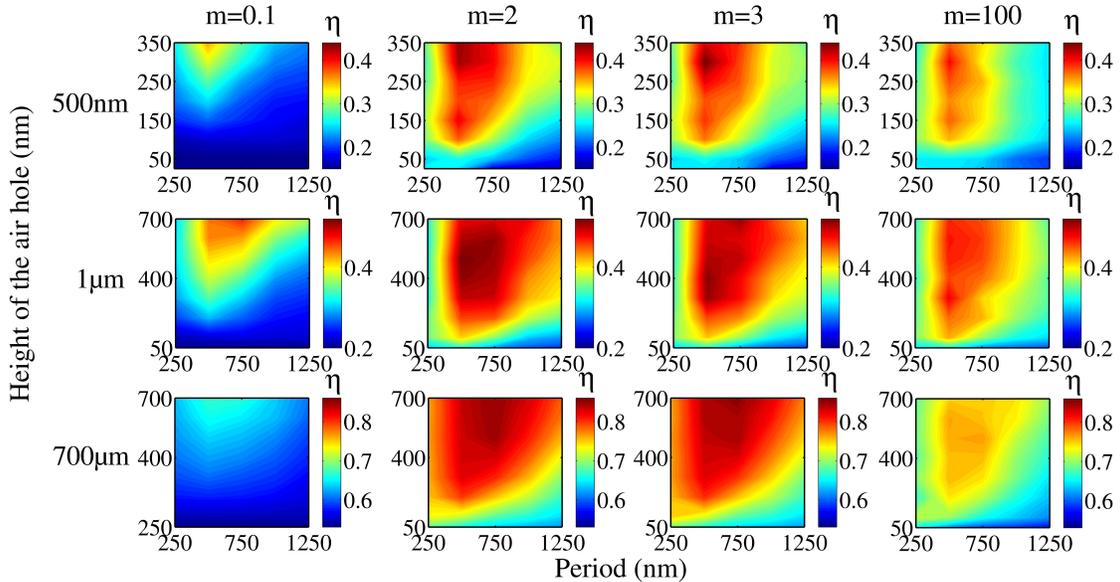

FIG. 3. (Color online) Maps of the integrated quantum efficiency ($\eta$) according to the period ($P$) and to the height of the holes ($h$) for different thicknesses and shapes of the air holes. The first row corresponds to an ultrathin slab of 500-nm thickness, the second row to an ultrathin slab of 1-µm thickness and the third row to a bulky slab of 700-µm thickness. The maps in the first column are for a constant hole shape of $m= 0.1$ (second column: $m=2$, third column: $m=3$, fourth column $m=100$). All these shapes are presented in Fig. 1.

From Fig. 3, we extract the optimal parameters for the three slab thicknesses. The optimal values are given in Table I. As it was shown by e.g. Deinega *et. al.*[27], we find that the optimal periods are of the order of the average between the free space wavelength and the wavelength inside silicon. As it was pointed out by Wellenzohn and Hainberger[45], we also observe that the period $P$ and the height $h$ of the holes both increase with the slab thickness $T$. This trend was already explained in terms of diffraction into the c-Si slab and interference between the diffracted orders.[15,16] In thicker cells, the period $P$ has to be increased in order to efficiently diffract low-energy light into the c-Si nanopatterned slab.[15] Regarding the height of the holes $h$, it was already shown that $h$ is responsible for interference between diffracted orders. Constructive interference between the diffracted orders corresponds to optimal heights.[16]

TABLE I. Optimal parameters according to the thickness of the c-Si slab.

|  | 500 nm c-Si slab | 1 µm c-Si slab | 700 µm c-Si slab |
|---|---|---|---|
| Period | 500 nm | 500 nm | 750 nm |
| Height of the hole | 300 nm | 500 nm | 700 nm |
| Shape | 3 | 2 | 3 |

For the different thicknesses of c-Si under study, let us now examine the influence of the shape on the merit factor $FOM_J$ (Fig. 4(a)) and on the photo-current $J$ (Fig. 4(b)).

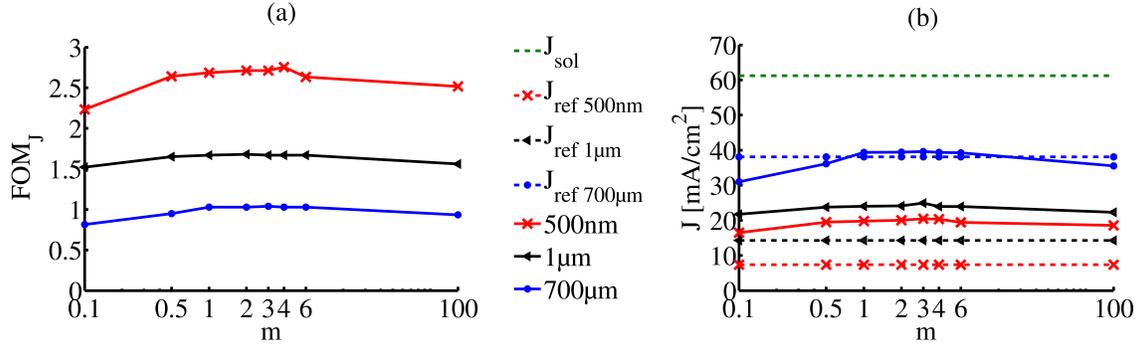

FIG. 4. (Color online) (a) Figure Of Merit of the photo-current ($FOM_J$) according to the shape of the holes for different thicknesses of c-Si (other parameters being optimal): 500 nm (crosses red line), 1 µm (triangles black line) and 700 µm (circles blue line). (b) Photo-current ($J$) according to the shape of the holes for different thicknesses of c-Si (all other parameters being optimal). The solar harvested photo-current ($J_{inc}$ - integrated over the entire solar spectrum) (green dashed line), the maximum achievable photo-current in the reference slabs of thicknesses 500 nm (crosses red dashed line), 1 µm (triangles black dashed line) and 700 µm (circles blue dashed line) are plotted as bench marks.

The thinner the c-Si slab, the higher the $FOM_J$, i.e. the more important are the nanostructures in the absorption enhancement (in comparison with the planar reference). As expected, the influence of nanopatterning is thus more important on ultrathin layers where absorption is smaller and where light-trapping schemes are critical (Fig. 4(a)).

Regarding the photo-current (Fig. 4(b)), we draw three major conclusions. First, we notice that the photo-current in the 500-nm patterned c-Si slab is always higher than the photo-current in the coated planar homogeneous slab of 1-µm thickness. This is very interesting since the nanostructured layer, in this case, has two times less absorbing material (500 nm versus 1 µm). Secondly, we notice the importance of the shape for the 700-µm bulky slab. Indeed, even with the best period, radius and height of the holes, choosing an inappropriate shape (*m<1* or *m>6*) leads to a lower photo-current than in the planar reference. Thirdly, although the shape of the holes has little influence on the maximum value of the photocurrent (for the 500-nm and 1-µm-thick slabs), it has a great importance on the size of the optimal region (Fig. 3). It is therefore interesting to enable tuning of the shape of the holes in the fabrication process. Indeed, the wider the optimal region, the higher the tolerance on experimental innacuracies in the parameter values.

## B. Discussion: Effect of the impedance matching and of the photonic light trapping

Absorption enhancement in the nanopatterned c-Si slabs is due to two processes: impedance matching and photonic light trapping (mode coupling). In order to characterize the impact of the shape of the holes on both mechanisms, we first examined the absorption spectra for different hole shapes (Fig. 5). Secondly, in order to identify the importance of both effects, we used the effective medium (EM) theory (Fig. 6). Two EM approximations were made (cfr. §III B). We focused on the 1-µm-thick c-Si layer patterned with the optimal parameters.

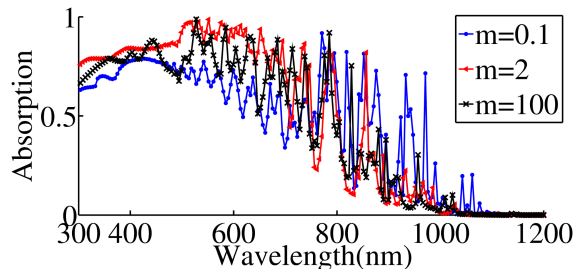

FIG. 5. (Color online) Absorption in a 1-µm-thick nanopatterned c-Si slab for different shapes of the hole (other parameters being optimal): *m=0.1* (circles blue line), *m=2* (triangles red line) and *m=100* (crosses black line).

In Fig. 5, we notice that the shape of the holes strongly influences the absorption in the c-Si slab. Changing the shape of the holes affects the absorption at short wavelengths (below 450 nm) where the enhancement is mainly due to impedance matching and at long wavelengths (above 450 nm) due to mode coupling, as it will be explained hereafter.

In the first use of the effective medium theory, we replaced the patterned part of c-Si by a single effective layer. In comparison with a planar homogeneous slab, we notice that the absorption is enhanced at short wavelengths (below 450 nm) (compare blue and magenta curves on Fig. 6(a)). This increase is due to the adaptation of the permittivity between the two layers. This first EM approximation gives a rough indication on the role of the hole in the anti-reflection process. However, it is not sufficient to explain the entire absorption enhancement (red curve).

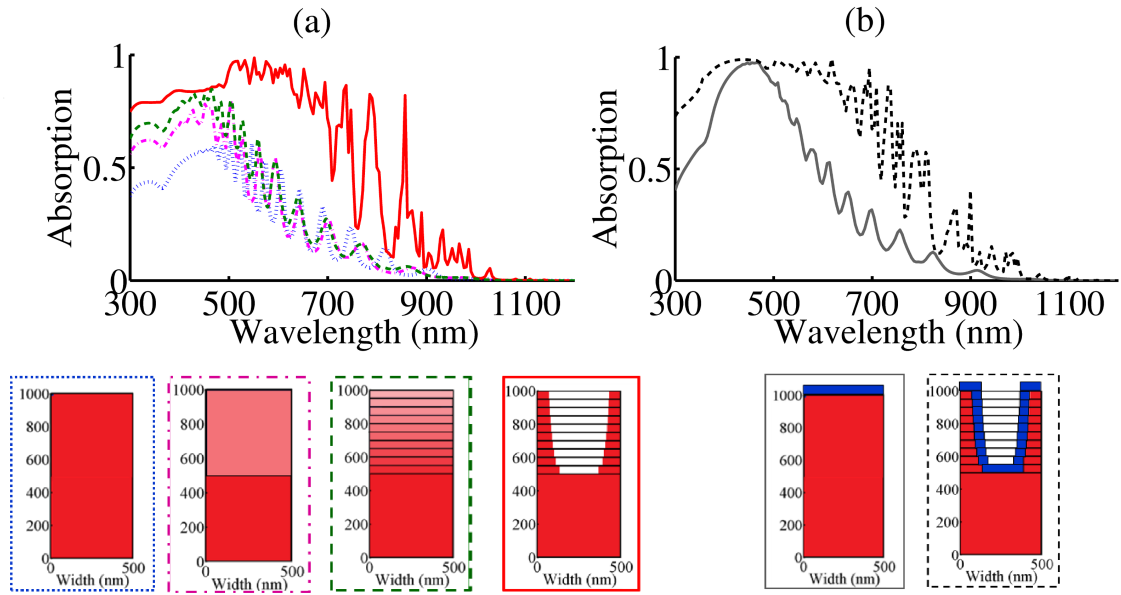

FIG. 6. (Color online) (a) Role of the holes in the anti-reflective effect and in the photonic light trapping. Absorption in a 1-µm-thick planar slab (dot blue line), the same slab with a single effective layer (dashed dot magenta curve), the same slab with several effective layers (graded index profile) (dashed green line) and the nanopatterned (optimal parameters) slab (solid red line). (b) Comparison between the absorption spectra of a 1-µm-thick coated (ARC=60 nm) slab (solid grey line) and a 1-µm-thick conformally coated (ARC=60 nm) nanopatterned (optimal parameters) slab (dashed black line).

Therefore, we used the effective medium theory a second time but we replaced the patterned part by several effective layers with same thicknesses as those of the layers. This method allowed us to identify the importance of the shape of the holes in the impedance matching. Indeed, by replacing the single effective layer by several effective layers we took the graded index profile (which is created by the hole) into account. As expected, we observe a further increase in the absorption (due to the graded index) at short wavelengths (below 450 nm) (compare the magenta and the green curves on Fig. 6 (a)).[18,46-48] Therefore, the holes are responsible for an anti-reflective effect (below 450 nm), in which the shape plays an important role due to the fact that it creates a graded index. Our structure allows a graded index effect, better than cylindrical holes and much better than anti-reflective coatings (single or double).

The absorption enhancement occurring in the absorption spectrum of the nanopatterned (optimal parameters) slab at longer wavelengths (above 450 nm) is not explained by the effective medium theory (compare the green and the red curves on

Fig. 6(a)). Hence, this increase is not due to an anti-reflective effect. However, Fig. 4 shows that thinner slabs (where light reaches the back side) are more improved than thick ones (where only the front side plays a role). So we deduce that back reflection inside the thin layer plays an important role (as it is the main difference between the thick and the thin slabs). This back reflection increases the path of the light inside the structure and therefore induces photonic light trapping.

In order to be more relevant for the comparison between the nanopatterned slab and the reference (ARC coated planar c-Si slab), we also calculated the absorption spectrum of a conformally coated (optimized thickness) nanopatterned 1-µm-thick c-Si slab (Fig. 6(b)). The absorption spectrum of the conformally coated, nanopatterned, slab exhibits two regions. In the first one, at wavelengths below 450 nm, the spectrum is more or less smooth, due to the fact that the absorption coefficient of c-Si is high, and therefore, light is efficiently absorbed. At these short wavelengths, the absorption enhancement (in comparison with the unpatterned coated slab) is due to an anti-reflective effect (graded index) depending on the shape of the holes. In the second region, at wavelengths above 450 nm, we observe light trapping due to the photonic structure. Light trapping is characterized by several intense peaks due to mode coupling. The peaks are smoother in the homogeneous slab, where mode coupling is not possible and are due, in this case, to Fabry Perot interferences between the flat interfaces at the front and back sides. These two regions also appear in amorphous silicon thin layers as it has been shown by Gomard *et al.*[49] and already explained.[13,35,50] It should be noted that this behavior is less pronounced in the 700-µm-thick c-Si slab (not shown here) due to the fact that an unpatterned slab already absorbs a big part of the incoming light. However, it also appears but the limit between the two regions is below 640 nm (for the anti-reflective effect) and above 750 nm (for the light-trapping).

## V. COMPARISON WITH A REAL NANOPATTERNED C-SI LAYER

To validate our model, we compare the simulated absorption spectra with experimental ones (Fig. 7(a-b)). Absorption spectra of the samples were obtained following the procedure described in section III D. It should be noted that, due to experimental constraints, the incidence angle was equal to 7°. The average pattern parameters of the samples were determined from Scanning Electron Microscopy (SEM) images (Fig. 7(c-f)). Some disorder is present on the patterned samples due to topographic inhomogeneities originating from experimental constrains of the NIL process. To take the disorder into account, several structures were simulated with periods and diameter varying by ±10% around average values. Averaging of the different simulations was performed leading to the final absorption spectrum. For both 1-µm-thick and 700-µm-thick nanopatterned c-Si slabs, the experimental and average simulated spectra were quasi identical (Fig. 7(a-b)).

It should be noted that for the simulations of the 1-µm-thick slab, the c-Si slab is not stand-alone in air like in the previous ones. Indeed, the real sample was bonded on glass (for handling reasons) and this was taken into account for the simulations. The wavelength-dependent complex permittivity for the glass was obtained from ellipsometric measurements.

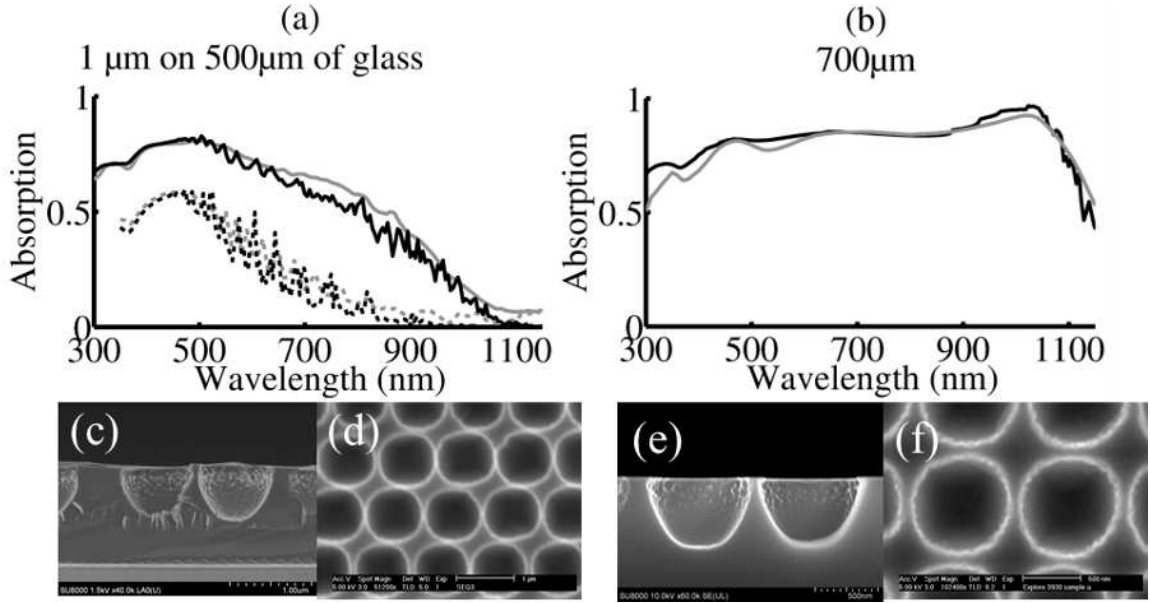

FIG. 7 (Color online) (a) Experimental absorption spectra (grey) and averaged simulated spectra (black) for a 1-µm c-Si nanopatterned uncoated film on a 500-µm glass substrate (solid lines) and for an unpatterned and uncoated film (dashed line). (b) Experimental absorption spectra (grey) and averaged simulated spectra (black) for a 700-µm c-Si nanopatterned uncoated film. (c-f) Scanning electron microscope (SEM) images of the 1-µm sample cross-section view (c) and top-view (d) and for the 700-µm sample cross-section view (e) and top-view (f) The averaged values for the 1-µm slab are: $h=670$ nm, $d=820$ nm, $m=1.25$, $P=900$ nm ($x$ axis) and *870 nm* (*y* axis for the 1-µm). The averaged values for the 700-µm slab are: $h=550$ nm, $d=795$ nm, $m=1.35$, $P= 930$ nm ($x$ axis) and *880 nm* (*y* axis). All the numerically simulated absorption spectra (black) were obtained by averaging a set of simulations performed with periods and diameters ranging from -10% to +10% around their average value. For all the numerically simulated spectra, we used unpolarized light with azimuthal and polar incidence angles equal to 7°.

## VI. CONCLUSIONS

Patterning with holes the front surface of a stand-alone c-Si slab increased the efficiency in comparison with a planar slab of the same thickness incorporating an anti-reflective coating (ARC). The use of a super-Gaussian function to describe the mathematical shape of the holes allowed us to investigate a large range of pattern morphologies and to study their influence on the light-harvesting efficiency of a solar cell.

We noticed that the optimal period and height highly depended on the thickness of the c-Si slab. On the contrary, independently of the thickness of the slab, the optimal diameter of the hole should be chosen as close as possible to the period. The use of non-cylindrical holes created a gradient of refractive index, which increased the absorption at short wavelengths (below 450 nm) via an anti-reflective effect. The holes also enhanced the absorption of wavelengths above 450 nm via diffraction giving a light trapping effect.

An optimally patterned and AR-coated 500-nm c-Si slab produced more than 1.4 times greater integrated photocurrent than an AR-coated 1-µm c-Si slab. It is therefore possible to increase the efficiency of ultrathin solar cells while reducing the amount of material by a factor of two.

The shape of the holes influenced the size of the optimal parameter region. This conclusion is of high importance for further technological applications. Indeed, it is more convenient to choose parameters in the broadest optimal zone. Hence, experimental inaccuracies (on the period and height of the holes) would have much less influence on the efficiency. If the shape of the holes is not suitably chosen, it is likely to cause the opposite effect: a decrease in the integrated photocurrent in comparison with a planar AR-coated slab.

The method we propose here could be helpful for the design and fabrication of ultrathin c-Si solar cells. Indeed, nowadays, front-side texturing techniques (such as random pyramid texturing) consume a lot of photoactive material. Therefore, efficient advanced light trapping consuming less quantity of material is needed. Ultrathin (500 nm) nanopatterned photoactive slabs, such as the one we propose in this article, could be the solution. Indeed, they reduce the cost of material and result in as high efficiencies as in a planar AR-coated slab of double thickness (1µm).

## ACKNOWLEDGMENTS

This research used resources of the Interuniversity Scientific Computing Facility located at the University of Namur, Belgium, which is supported by the F.R.S.-FNRS under convention No. 2.4617.07.